\documentclass[12pt]{article}
\usepackage{epsfig}
\usepackage{amsfonts}
\usepackage{amsopn}
\usepackage{amsmath}

\newcommand{\eq}{\begin{equation}}
\newcommand{\eqx}{\end{equation}}
\newcommand{\eqn}{\begin{eqnarray}}
\newcommand{\eqnx}{\end{eqnarray}}
\newcommand{\f}[2]{\frac{#1}{#2}}

\newcommand{\dl}{\delta}
\newcommand{\lm}{\lambda}
\newcommand{\nn}{{\cal N}}
\newcommand{\bps}{$\f{1}{16}$BPS }

\DeclareMathOperator{\tr}{tr}

\newcommand{\nai}{{n_{a_1}}}
\newcommand{\naii}{{n_{a_2}}}
\newcommand{\nbi}{{n_{b_1}}}
\newcommand{\nbii}{{n_{b_2}}}
\newcommand{\nci}{{n_{c_1}}}
\newcommand{\ncii}{{n_{c_2}}}
\newcommand{\nciii}{{n_{c_3}}}
\newcommand{\nciv}{{n_{c_4}}}

\newcommand{\na}{n_a}
\newcommand{\nb}{n_b}
\newcommand{\nc}{n_c}

\newcommand{\qqqq}{\quad\quad\quad\quad}

\title{Supergravitons from one loop perturbative $\nn=4$ SYM}

\author{
Romuald A. Janik$^a$ \thanks{e-mail: {\tt  ufrjanik@if.uj.edu.pl}} \
and Maciej Trzetrzelewski$^{a,b}$ \thanks{e-mail: {\tt
    33lewski@th.if.uj.edu.pl}} \\ \\ 
$^a$ Institute of Physics,\\
Jagiellonian University, \\
Reymonta 4, 30-059 Krak\'ow,\\
Poland \\ \\
$^b$  Department of Mathematics,\\
Royal Institute of Technology, \\
KTH 407.76, 100-44 Stockholm, \\
Sweden.
}

\begin{document}

\maketitle

\begin{abstract}
We determine the partition function of \bps operators in $\nn=4$ SYM
at weak coupling at the one-loop level in the planar limit. This
partition function is significantly different from the one computed at
zero coupling. We find that it coincides precisely with the partition
function of a gas of \bps `supergravitons' in $AdS_5 \times S^5$. 
\end{abstract}

\vfill

\section{Introduction}

The AdS/CFT correspondence states an exact equivalence between $\nn=4$
SYM gauge theory and type IIB superstrings in an $AdS_5 \times S^5$
background \cite{adscft}. It provides a fascinating new approach for
studying nonperturbative properties of gauge theory. On the other
hand, one can use the gauge theory knowledge to gain insight into the
behaviour of (super-)gravity at the quantum level
(see e.g. \cite{berenstein,mandal,maoz}). In general this is a
formidable problem but 
progress can be made when studying configurations which preserve some
fraction of supersymmetry. A dictionary between $\f{1}{2}$BPS
operators in gauge theory and dual geometries has been established in
\cite{LLM}. $\f{1}{4}$- and $\f{1}{8}$BPS states have been discussed
from various points of view \cite{other}. Of particular interest are the \bps
states \cite{MM} due to the existence \bps black holes
\cite{bpsbh}. At low energies, the gauge theory \bps states should
correspond to a gas of 
\bps supergravitons, while at high energies these states should
account for the entropy of \bps black holes. 

In \cite{MM} \bps states were counted on the gauge theory side at
zero coupling. It was found that the resulting partition function   
overcounts both the \bps supergraviton partition function (giving a
different energy scaling of entropy) and the \bps black hole entropy
in the relevant parameter regimes. In that paper it was suggested that
once gauge theory interactions are turned on, many states which were
counted as \bps at zero coupling would get anomalous dimensions, and
that the overcounting could be cured.

The aim of this paper is to perform the enumeration of \bps operators in
perturbative $\nn=4$ SYM to one-loop order. We do
the counting in the planar limit using the oscillator construction of
the one-loop dilatation operator of \cite{Beisert}. We find {\em
  exact} agreement with the \bps supergraviton partition function.

The plan of the paper is as follows. In section 2 we review the
definition of \bps states and fix notation. Then, in section 3, we
review the counting of these states in the free theory, and in section
4 we describe what has to be done to perform the calculation at one
loop. In section 5 we review the supergravity result for the \bps
supergraviton partition function. In section 6 we describe in some
details the construction of the one loop dilatation operator and, in
the following section, we determine the partition function of \bps
operators and perform some checks. In section 8 we compare the result
with the supergravity prediction and finaly, in section 9, we discuss the
possible extension to large but finite $N$. We close the paper with a
summary. 

\section{\bps states}

In this paper we consider \bps states which by definition are
annihilated by the following two supercharges: 

\eq 
Q \equiv Q^{-\f{1}{2},1}, \qqqq S \equiv S^{-\f{1}{2},1}, 
\eqx 
where $Q^\dagger=S$ and $Q^{-\f{1}{2},1}$, $S^{-\f{1}{2},1}$ are as in
\cite{MM}. We would like to calculate the partition function over
these states. To do so it is convinient to introduce the anticommutator 

\eq \Delta \equiv 2\{ S,Q \}. \eqx 
The states annihilated by $S$ and $Q$
are exactly those annihilated by $\Delta$. Moreover these states are
in a 1 to 1 correspondence with the cohomology classes w.r.t. $Q$.

In general, states in $\nn=4$ SYM can be labeled by the eigenvalue
of the dilatation operator $H$, two Lorentz spins $J_1$ and $J_2$,
and three $SU(4)_R$ charges $R_1$, $R_2$ and $R_3$ (we use the
notation of \cite{MM}). The anticommutator $\Delta$ can be evaluated
in terms of these quantum numbers. We have 

\eq \Delta =2 \{ S,Q\}=H-2J_1-\f{3}{2}R_1 -R_2 -\f{1}{2} R_3. \eqx 
Hence we have to
calculate the partition function 
\eq 
Z_{\f{1}{16}BPS}=\tr_{\Delta=0}x^{2H} z^{2J_1} y^{2J_2} v^{R_2}
w^{R_3}, \eqx  
where we picked just
one possible choice of generating parameters. $R_1$ is of course
fixed by the condition $\Delta=0$. The above partition function
includes all single and multitrace operators. It counts {\em all}
operators annihilated by $Q$ and $S$. We thus do not restrict
ourselves to operators which are \bps but not $\f{1}{8}$BPS or
higher.

\section{Gauge theory at zero coupling}

In order to evaluate the partition function in gauge theory it is
convenient to use the oscillator representation, introduced in
\cite{Beisert}, for all single trace 
operators. In this picture, a single
trace operator $\tr O_1O_2O_3\ldots O_L$ is represented by $L$ sites
each occupied by the `elementary' field $O_i$. Operators $O_i$ are in turn
represented by states in a Fock space generated by 4 bosonic
($a_1^\dagger,a_2^\dagger$ and $b_1^\dagger,b_2^\dagger$) creation
operators, and 4 fermionic ones ($c^\dagger_1, c^\dagger_2
,c^\dagger_3, c^\dagger_4$). The Fock space is narrowed by the
central charge constraint which relates the total number of
oscillators of various kinds on each site:

\eq
\label{e.cc}
\na-\nb+\nc=2.
\eqx
For an explicit dictionary between operators and Fock space states see
\cite{Beisert}.

The Lorentz spins $J_1,J_2$ and the
$SU(4)$ charges are simply represented by the total number of various
oscillators\footnote{The sum $\sum_{i=1}^3 q_i$ is defined as
  $\f{3}{2}R_1+R_2+\f{1}{2}R_3$.}:

\eqn
\label{e.chargedef}
R_1 &=& \ncii-\nci \nonumber,\\
R_2 &=& \nciii-\ncii \nonumber,\\
R_3 &=& \nciv-\nciii \nonumber,\\
\sum_{i=1}^3 q_i &=& \f{1}{2} (\nci+\ncii+\nciii+\nciv)-2\nci \nonumber,\\
J_1 &=& \f{1}{2} (\naii-\nai) \nonumber,\\
J_2 &=& \f{1}{2}( \nbii-\nbi). 
\eqnx 
In the {\em free theory}, the
free dilatation operator $H_0$ also has a similar representation 

\eq \label{e.h0} H_0= \nai+\naii+\f{1}{2}(\nci+\ncii+\nciii+\nciv). \eqx
Consequently, in the free SYM theory, the condition $\Delta=0$ can
be evaluated to give 

\eq \Delta_{\lambda=0}= 2\nai+2\nci =0. \eqx 
Therefore, \bps states are exactly the operators which do not have any
$a_1^\dagger$ 
or $c_1^\dagger$ operators in the oscillator representation. Since
all the spins and charges are expressed in terms of the total number
of oscillators of various kinds, it is convenient to keep track of
the number of oscillators of each type when counting \bps operators.
We thus consider partition functions of the form 

\eq Z(a_2,b_1,b_2,c_2,c_3,c_4) = \sum_{\Delta=0} a_2^\naii b_1^\nbi
b_2^\nbii c_2^\ncii c_3^\nciii c_4^\nciv . \eqx
A simple counting over the Fock space states, taking into
consideration the central 
charge constraint (\ref{e.cc}), gives for the `letter' partition
function (partition function of operators at each site): 

\eqn
\label{e.letter}
z_B &=& \f{a_2^2+c_2 c_3+c_2c_4+c_3c_4}{(1-b_1 a_2)(1-b_2 a_2)},\\
z_F &=& \f{a_2(c_2+c_3+c_4)+(b_1+b_2-a_2b_1b_2)c_2c_3c_4}{(1-b_1 a_2)
  (1-b_2 a_2)},
\eqnx
where we made a separation into bosonic and fermionic states.

The partition function of single trace operators then follows from

\eq
\label{e.zstfree}
Z_{s.t.}=-\sum_{n=1}^\infty \f{\phi(n)}{n} \log \left(
1-z_B(x^n)-(-1)^{n+1} z_F(x^n) \right) ,
\eqx
where $x$ stands for generic arguments
(e.g. $x=(a_2,b_1,b_2,c_2,c_3,c_4)$ in our case).
Finally, the partition function of multitrace operators (at
$N_c=\infty$) is given by

\eq
\label{e.zninf}
Z=\exp \left( \sum_{n=1}^\infty \frac{1}{n} \left\{Z_{s.t.}^B(x^n)
+(-1)^{n+1} Z_{s.t.}^F(x^n) \right\} \right).
\eqx
The above formulas do not take into account finite $N$ effects which
appear, e.g. when certain long traces are equivalent to linear
combinations of shorter multitrace operators (due to the
Cayley-Hamilton theorem for finite matrices). For the specific case
of {\em free} SYM, the method of character expansions of
\cite{Sundborg,Aharony} allows to perform an exact calculation at
finite $N$ starting directly from the letter partition function
(\ref{e.letter}). The resulting fixed $N$ partition function is
given by the formula 

\eq \label{e.fixednfree} Z=\int DU \exp \left\{
\sum_{n=1}^\infty \left(z_B(x^n)+(-1)^{n+1} z_F(x^n)\right) \f{\tr
U^n \tr U^{-n}}{n} \right\}, \eqx 
where the integral is over the
unitary group $U(N)$. For the case at hand this has been analyzed in
\cite{MM} for large $N$. For small values of parameters $x$ the
large $N$ limit of (\ref{e.fixednfree}) does not depend on $N$
(reproducing effectively 
(\ref{e.zninf})), while at a finite value of $x$ (strictly less than
1) the \bps partition function exhibits a behavior $\log Z \sim
N^2$.

\section{Gauge theory at one loop}

At one loop, $H=H_0 + \lambda  \dl H$ and the anomalous part is now a
nontrivial operator which acts on each two neighboring sites. The
complete one loop dilatation operator was constructed in
\cite{Beisert}. We discuss it in details in section 6 . The
condition $\Delta_{1-loop}=0$ now takes the form 

\eq
\Delta_{1-loop}=\Delta_{\lm=0}+\lm \dl H=0. \eqx 
Since at one loop the
eigenvalues of $\dl H$ are rational/radical expressions, for generic
transcendental $\lm$ this condition picks out states which satisfy
{\em both} the free and one loop conditions separately, i.e. states
with $\nai=\nci=0$ which do not get any anomalous dimensions $\dl H$
at one loop. We thus have to compute 

\eq
Z=\sum_{\stackrel{\nai=\nci=0}{\dl H=0}} a_2^\naii b_1^\nbi
b_2^\nbii c_2^\ncii c_3^\nciii c_4^\nciv . \eqx 
Note that now the
formula (\ref{e.zstfree}) no longer holds and we have to identify the number of
operators which do not get anomalous dimensions at one loop for each
$L$ independently.

We first determine the sum over single trace operators for fixed $L$ by
computing the above power series with some truncation on the number
of oscillators. This turns out to give enough information to guess
the analytical form of the generating function. Next, we test the
function on various configurations which were not used in the
process of obtaining the analytical form. The details of this
procedure are discussed in section 7.

Summing over $L$ gives the partition function of \emph{all} single trace
operators. Then (\ref{e.zninf}) may be used to get the partition
function of multitrace ones. Let us note that at one loop we do not
have a counterpart of the exact formula for finite $N$ valid for
zero coupling (\ref{e.fixednfree}). We will, nevertheless, discuss
some aspects of the possible large $N$ behavior at the end of the
paper.

\section{Supergraviton partition function}

At strong coupling one can calculate the partition
function over \bps states using the supergravity/superstring side of
the AdS/CFT correspondence. This has been considered in \cite{MM}.
We now briefly review these results.

Since the $psu(2,2|4)$ supersymmetry algebra of the gauge theory is
also the symmetry group of superstrings in $AdS_5 \times S^5$, we have
direct counterparts of $Q$ and $S$ operators and we can use them
to define the \bps states.

In the low energy regime the partition function should be given by
supergravity fields which are annihilated by the $Q$ and $S$
operators. This has been done in \cite{MM} where the single particle
partition function 

\eq Z^{single}_{gravitons}=\sum_{\Delta=0} x^{2H}
z^{2J_1} y^{2J_2} v^{R_2} w^{R_3}, \eqx 
was calculated with the result 

\eqn \label{e.sugra}
Z^{single}_{gravitons} &=& \f{bosons+fermions}{denominator}, \\
denominator &=& (1-\f{x^2}{w})(1-x^2 v)(1-x^2 \f{w}{v})(1-x^2\f{z}{y})
(1-x^2 zy), \\ 
bosons &=& v x^2+\f{x^2}{w} +\f{w x^2}{v} -\f{x^4}{v} -\f{v x^4}{w} -w
x^4 +2 x^6 + \f{x^6 z}{y v} \nonumber\\
&&  +\f{v x^6 z}{w y} +\f{w x^6 z}{y} -\f{x^8
  z}{y} +\f{x^6 z y}{v}+\f{v x^6 z y}{w} +w x^6 z y \nonumber\\
&&-x^8 z y+x^4 z^2+x^{10} z^2, \\
fermions &=& \f{x^3}{y}+x^3 y+\f{x^3 z}{v}+\f{v x^3 z}{w} +w x^3 z-2
x^5 z+ v x^7 z \nonumber\\
&& + \f{x^7 z}{w}+\f{w x^7 z}{v} +\f{x^7 z^2}{y} +x^7 z^2 y.
\eqnx
The full partition function is obtained by summation over the Fock
spaces of these particles using the formula

\eq
\label{e.sugramulti}
Z_{gravitons}=\exp \left( \sum_{n=1}^\infty
\frac{1}{n} \left\{ Z^{single,bos}_{gravitons}(x^n,\ldots,w^n)
+(-1)^{n+1} Z^{single,fer}_{gravitons}(x^n,\ldots,w^n) \right\} \right).
\eqx
We note that the above formula is identical in form to the one
obtained when passing from single- to multi-trace operators in gauge
theory (\ref{e.zninf}).

It turns out that $Z_{gravitons}$ {\em does not} agree with the
result from free SYM theory \cite{MM}. In the case when $z=y=v=w=1$ even the
scaling of the entropy with energy is different.

Moreover, \cite{MM} obtained the partition function for $\f{1}{8}$BPS
states by taking the limit $z \to 0$. The result again was in
disagreement with free SYM, but matched exactly the calculation made
using properties of the chiral ring of (interacting) SYM.

In section 8 we compare this supergraviton partition
function with {\em perturbative} computations at one loop in SYM.

When energies are large (compared with $N$) it is expected that the
partition function for \bps states will have a 

\eq \log Z \sim N^2,\eqx 
behavior which should coincide with the one obtained from the \bps
black holes (see \cite{MM}, section 5.3 for explicit formulas). In \cite{MM}, a
similar qualitative behavior was obtained at zero coupling, although
the numerical details did not match. The motivation for this paper
was to investigate how much of this zero coupling result survives at
one loop.

\section{The one loop dilatation operator}

In this section we review the construction of the one loop
dilatation operator $\dl H$ in the oscillator picture  \cite{Beisert}.

\subsubsection*{The Fock space}

Let us consider the space of operators which are traces of $L$
adjoint fields. We represent each of those fields as a state on
one of $L$ sites. Since the trace is cyclic invariant we restrict
ourselves to cyclic invariant states 
of the Fock space at the end of
the calculation.

A generic state in Fock space is thus a linear combination of states

\eq \mid s_1 \rangle \otimes \ldots  \otimes \mid s_L \rangle,
\label{basis} \eqx 
where on each site $i$ the
state $\mid s_i \rangle$ is obtained by acting with bosonic
$a_{1,i}^{\dagger},a_{2,i}^{\dagger}, b_{1,i}^{\dagger},
b_{2,i}^{\dagger}$ and fermionic $c_{1,i}^{\dagger},
c_{2,i}^{\dagger}, c_{3,i}^{\dagger}, c_{4,i}^{\dagger}$ creation
operators on the Fock vacuum $\mid 0 \rangle_i$. An arbitrary state
is labeled by the oscillator occupation numbers 

\eqn \mid s_i
\rangle &=&\mid n_{a_{1,i}},n_{a_{2,i}},n_{b_{1,i}},n_{b_{2,i}},
n_{c_{1,i}},
n_{c_{2,i}}, n_{c_{3,i}},n_{c_{4,i}}\rangle \nonumber\\
&=& a_{1,i}^{\dagger \ n_{a_{1,i}}} a_{2,i}^{\dagger \ n_{a_{2,i}}}
b_{1,i}^{\dagger \  n_{b_{1,i}}} b_{2,i}^{\dagger \ n_{b_{2,i}}}
c_{1,i}^{\dagger \ n_{c_{1,i}}} c_{2,i}^{\dagger \ n_{c_{2,i}}}
c_{3,i}^{\dagger \ n_{c_{3,i}}} c_{4,i}^{\dagger \ n_{c_{4,i}}} \mid
0 \rangle_i . \eqnx 
As discussed in section 3, the occupation
numbers at each site are constrained by (\ref{e.cc}) 

\eq
n_{a_{2,i}}+n_{a_{2,i}}-n_{b_{1,i}}-n_{b_{2,i}}+ n_{c_{1,i}}+
n_{c_{2,i}}+ n_{c_{3,i}}+n_{c_{4,i}}=2. \eqx 
The one loop dilatation
operator does not change the total number of oscillators of each
kind and only moves them from site to site. Therefore it acts within
the space with fixed total number of oscillators of any type for
fixed $L$. It follows that we can diagonalize $\dl H$ in subspaces
labeled by 

\eq [\nai,\naii,\nbi,\nbii,\nci,\ncii,\nciii,\nciv;L].
\eqx

\subsubsection*{The harmonic action}

Let us now review the construction of $\dl H$ \cite{Beisert},
giving more details about the computer code implementation.

The action of the one-loop dilatation operator introduces an
interaction between \emph{only} the neighboring sites (the last and the
first site are assumed to be neighbors). For this reason it is
enough to consider a pair of such sites
\begin{equation}
\mid v\rangle=\mid
n_{a_1},n_{a_2},n_{b_1},n_{b_2},n_{c_1},n_{c_2},n_{c_3},n_{c_4}\rangle
\otimes \mid
m_{a_1},m_{a_2},m_{b_1},m_{b_2},m_{c_1},m_{c_2},m_{c_3},m_{c_4}\rangle
\label{1el},
\end{equation}
where we dropped the index $i$ ( and $i+1$ ) for clarity. Our object
is now to calculate the hamiltonian matrix element between
(\ref{1el}) and an arbitrary other state
\begin{equation}
\mid v'\rangle=\mid
n'_{a_1},n'_{a_2},n'_{b_1},n'_{b_2},n'_{c_1},n'_{c_2},n'_{c_3},n'_{c_4}\rangle
\otimes \mid
m'_{a_1},m'_{a_2},m'_{b_1},m'_{b_2},m'_{c_1},m'_{c_2},m'_{c_3},m'_{c_4}\rangle
\label{2el}.
\end{equation}
If one is interested in calculating the element $\langle v' \mid
H \mid v \rangle$  then  it turns out that the harmonic action \cite{Beisert}
can be described by the following set of rules
\begin{itemize}

\item  consider all the possibilities of oscillator hopping from site
  $i \to i+1$ and from site $i+1 \to i$ such that the state
  (\ref{1el}) becomes (\ref{2el})

\item to each such possibility associate a number
\eq
c_{n,n_{12},n_{21}}=(-1)^{1+n_{12}n_{21}}\frac{\Gamma(\frac{1}{2}n_{12}+
  \frac{1}{2}n_{21})
\Gamma(1+\frac{1}{2}n-\frac{1}{2}n_{12}-\frac{1}{2}n_{21})}{\Gamma(1+
  \frac{1}{2}n)},
\eqx
where $n_{12}$, $n_{21}$ are the numbers of oscillators hopping from
$i \to i+1$, $i+1 \to i$ respectively, $n$ is the total number of
quanta at sites $i$ and $i+1$ in the beginning
\item include the $-1$ factors when the fermion oscillators are
  hopping "over" other fermions.
In particular, if a fermion is hopping form $i=1$ to $i=L$ or vice
versa then all fermions in between (for $1<i<L$) have to be
considered.

\item sum over all the possibilities and multiply the result by $
  \frac{\left\| \mid v' \rangle \right\|}{ \left\| \mid v \rangle
    \right\|}$

\end{itemize}

The harmonic action can be implemented in two independent ways. One is
to use the above rules as they are and compute the element $\langle v'
\mid H \mid v \rangle$ indirectly by evaluating
\eq
H\mid v \rangle = \sum_{v'}H_{v,v'}\mid v' \rangle.
\eqx
Second is to write down the formula for the matrix element $\langle v'
\mid H \mid v \rangle$ and compute it explicitly. It turns out to be
possible, we have

\[
\langle v' \mid H \mid v \rangle =\sqrt{ \prod_{t \in T}\frac{
    {n'}_t!}{n_t!}}  \sum_{t' \in T}\sum_{k_{t'}=0}^{m_{t'}}
(-1)^Fc_{x,y,z} \prod_{t'' \in T}
\binom{n_{t''}}{n_{t''}-n'_{t''}+k_{t''}}\binom{m_{t''}}{k_{t''}}, 
\label{matel} 
\]
\[
T=\{a_1,a_2,b_1,b_2,c_1,c_2,c_3,c_4\},
\]
\begin{equation}
x=\sum_{t \in T} n_t+m_t, \ \ \ \ y=\sum_{t \in T}
\epsilon_{t}|n_{t}-n'_{t}|+ k_{t}, \ \ \ \ z=\sum_{t \in T}
\eta_{t}|n_{t}-n'_{t}|+k_{t}, 
\end{equation}
where $(-1)^F$ is the fermion number discussed in one of the rules and
where the parameters $\epsilon_t$, $\eta_t$ are defined in the
following way. 
$\epsilon_t$ is equal $1$  if $n_t \ge n'_t$ and $0$ otherwise,
$\eta_t=1-\epsilon_t$.

The above formula can be justified in the following way. Let us
consider $n_{a_1}$ bosons $a_1^{\dagger}$. We want them to hop form
$i\to i+1$ so that only $n'_{a_1}$ of them are left ( clearly we
assume that $n_{a_1} \ge n'_{a_1} $). Since they commute and are
indistinguishable the number of possibilities coincides with the
number of combinations $\binom{n_{a_1}}{n_{a_1}-n'_{a_1}}$. Other
possibilities are when the number of such hops is
$n_{a_1}-n'_{a_1}+k_{a_1}$ with $k_{a_1}>0$. Then, we have to hop back (from
$i+1\to i$) exactly $k_{a_1}$ oscillators $a_1^{\dagger}$. This can be done in
$\binom{m_{a_1}}{k_{a_1}}$ ways. Therefore, the net factor for given
$k_{a_1}$ is
$\binom{n_{a_1}}{n_{a_1}-n'_{a_1}+k_{a_1}}\binom{m_{a_1}}{k_{a_1}}$. 
To include all the possibilities we sum over all possible $k_{a_1}$'s,
i.e from $0$ to $m_{a_1}$.

For the other bosonic and fermionic operators $a_2^{\dagger}, \
b_1^{\dagger}, \ b_2^{\dagger}, \ c_1^{\dagger}, \ c_2^{\dagger}, \
c_3^{\dagger}, \ c_4^{\dagger} $ the analysis in analogous and gives
the corresponding factors as in (\ref{matel}). To include the $-1$
factors coming from hopping of fermions we weight the sum
(\ref{matel}) with the factor $(-1)^F$.

We have implemented the above construction of the one loop dilatation
operator independently in two different programs and verified that the
results agree. As a further check we reproduced  various one loop
anomalous dimensions given in \cite{Beisert}.

In order to complete the construction we project the Hilbert space
(\ref{basis}) to the 
subspace of states which are invariant under cyclic permutations,
since only these states correspond to gauge theory single trace
operators.

We start with the hamiltonian matrix $H$ represented in the
\emph{non}-cyclic invariant basis $B$ (\ref{basis}) constructed as
above. Then, we
construct a matrix 
representation of an operator $T$ which translates the chain by one
site. The cyclic invariant states correspond to eigenvectors
$v_1,\ldots,v_n$, $\ n \le \#B$ of $T$ with an eigenvalue equal 1.
Now, we build the projection matrix
$P=[v_1,\ldots,v_n]$ and perform the similarity transformation
\[
H \to PHP^T,
\]
on $H$. The result is the hamiltonian matrix
represented in the cyclic invariant basis.

\section{The one loop \bps partition function}

According to the general discussion in previous sections, the tree
level condition for the \bps states contributing to the index is
$\Delta_{\lambda=0}=2n_{a_1}+2n_{c_1}=0$ hence from now on we take
$n_{a_{1,i}}=n_{c_{1,i}}=0$. Moreover, at one loop level the
condition $\Delta_{1-loop}=0$ is satisfied only for states which are
eigenstates corresponding to $0$ eigenvalue of the one loop dilatation
operator. Let
$D_{n_{a_2},n_{b_1},n_{b_2},n_{c_2},n_{c_3},n_{c_4},L}$ be the
number of such states in the sector with $n_{a_2}$, $n_{b_1}$,
$n_{b_2}$, $n_{c_2}$, $n_{c_3}$, $n_{c_4}$ number of quanta and $L$
sites respectively. The generating function we are looking for is
\[
Z_L^{1/16th}(a_2,b_1,b_2,c_2,c_3,c_4)= \hspace{-0.8cm}
\sum_{\stackrel{n_{a_2},n_{b_1},n_{b_2}=0,\ldots,\infty}{n_{c_2},n_{c_3},
    n_{c_4}=0,\ldots, L }} \hspace{-0.8cm}
D_{n_{a_2},n_{b_1},n_{b_2},n_{c_2},n_{c_3},n_{c_4},L} a_2^{n_{a_2}}
b_1^{n_{b_1}} b_2^{n_{b_2}} c_2^{n_{c_2}} c_3^{n_{c_3}}
c_4^{n_{c_4}},
 \]
(the sum over fermionic variables runs from $0$ to $L$ due to the
Pauli exclusion principle). With use of computer code implementation of 
the rules discussed in previous section, one can determine the
numbers $D_{n_{a_2},n_{b_1},n_{b_2},n_{c_2},n_{c_3},n_{c_4},L}$
exactly, but of course only for a finite number of configurations.
It is by no means obvious that such data can determine the whole
function\\ 
$Z_L^{1/16th}(a_2,b_1,b_2,c_2,c_3,c_4)$. Nevertheless, our analysis
shows that the Taylor expansion of the function
$Z_L^{1/16th}(a_2,b_1,b_2,c_2,c_3,c_4)$ coincides with the expansion of
certain rational function. The details of our computation are below.

For $L=2$ we analyzed the configuration with $0 \le n_{a_2}, \ n_{b_1}, \
n_{b_2} \le 10$ and $ 0 \le n_{c_2}, \ n_{c_3}, \ n_{c_4} \le
2$. There are $11^33^3=35937$ such possibilities however
only $1494$ of them satisfy the central charge constraint.

For $L=3$ we took $0 \le n_{a_2}, \ n_{b_1}, \ n_{b_2} \le 5$ and $ 0
\le n_{c_2}, \
n_{c_3}, \ n_{c_4} \le 3$. There are $6^3 4^3=13824$ such possibilities
among which only $849$ satisfy the central charge constraint.

For $L=4$ we analyzed the configuration with $0 \le n_{a_2}, \ n_{b_1}, \
n_{b_2} \le 2$ and $ 0 \le n_{c_2}, \ n_{c_3}, \ n_{c_4} \le 4$. There are $3^3
5^3=3375$ such possibilities and  $279$ which satisfy the central
charge constraint.

Let us now explain how the partition function was
reconstructed from the above data and consider in detail the case of
$L=2$. The computer analysis gives a polynomial
\[
Z_{L=2}^{1/16th, cut}(a_2,b_1,b_2,c_2,c_3,c_4)=
 \hspace{-0.8cm}
\sum_{\stackrel{n_{a_2},n_{b_1},n_{b_2}=0,\ldots, 10}{n_{c_2},n_{c_3},
    n_{c_4}=0,1,2 }} \hspace{-0.8cm}
D_{n_{a_2},n_{b_1},n_{b_2},n_{c_2},n_{c_3},n_{c_4},2} a_2^{n_{a_2}}
b_1^{n_{b_1}} b_2^{n_{b_2}} c_2^{n_{c_2}} c_3^{n_{c_3}}
c_4^{n_{c_4}},
 \]
which consists of $1494$ terms. 

Our strategy to proceed is the following. If
the full partition function $Z_{L=2}^{1/16th}(a_2,b_1,b_2,c_2,c_3,c_4)$ is a
rational function then, in particular, so is $Z_{L=2}^{1/16th}(a_2,1,1,1,1,1)$.
Therefore, the coefficients of $Z_{L=2}^{1/16th}(a_2,1,1,1,1,1)$ should be
"easily" recognizable. Indeed, we have
\eqn
&& Z_{L=2}^{1/16th, cut}(a_2,1,1,1,1,1) =13 + 40 a_2 + 72 a_2^2 + 104 a_2^3 +
    136 a_2^4 + 168 a_2^5  \nonumber \\
 && \qqqq +   200 a_2^6 + 232 a_2^7 + 264 a_2^8 + 296 a_2^9 + 320 a_2^{10},
\eqnx
which (except for the last term $320 a_2^{10}$) is recognized as the
Taylor expansion of
\[
\frac{13+14a_2+5a_2^2}{(1-a_2)^2}.
\]
Next, we turn on the variable $b_1$, i.e. we consider
$Z_{L=2}^{1/16th, cut}(a_2,b_1,1,1,1,1)$ and find the corresponding generating
function. Then, we proceed analogously with
$Z_{L=2}^{1/16th, cut}(a_2,b_1,b_2,1,1,1)$ and find that
\[
Z_{L=2}^{1/16th}(a_2,b_1,b_2,1,1,1)=\frac{6 + 8 a_2 + 3 a_2^2 + (3 + 3  a_2
  + a_2^2)(b_1 + b_2) + b_1 b_2}{(1-b_1 a_2)(1-b_2a_2)}.
\]
The full $a_2$, $b_1$, $b_2$ dependence is now determined. In order
to find the $c_2$, $c_3$, $c_4$ dependence we do the following.
First, due to the Pauli exclusion principle the fermionic variables
cannot be in the denominator $(1-b_1 a_2)(1-b_2a_2)$. Therefore they
enter only in the numerator in the form $c_2^ic_3^jc_3^k$, $i,j,k
\le L$. Second, the harmonic action is completely symmetric with
respect to fermionic oscillators. This implies that the numerator of
$Z_{L=2}^{1/16th}(a_2,b_1,b_2,c_2,c_3,c_4)$ is a completely symmetric function
with respect to $c_2$, $c_3$ and $c_4$. \footnote{By the same
argument the full partition function has to be symmetric in bosonic
variables $b_1$, $b_2$ hence only the combinations $b_1+b_2$, $b_1
b_2$ appear in the numerator. It is not obvious why there are no
other, higher order, combinations e.g. $b_1^3+b_2^3$. Clearly, this
must be a property of the harmonic action. }. We therefore write
$Z_{L=2}^{1/16th}(a_2,b_1,b_2,c_2,c_3,c_4)$ as
\begin{equation}
\frac{1}{(1-b_1
  a_2)(1-b_2a_2)}\sum_{n=0}^2\sum_{m=0}^2\sum_{l_1,l_2,l_3=0}^2
\tilde{D}_{n,m,l_1,l_2,l_3 }A_n B_m \sigma_{l_1,l_2,l_3}(c_2,c_3,c_4),
\label{shur}
\end{equation}
where
\[
A_0=1, \ A_1=a_2, \ A_2=a_2^2,
\]
\[
B_0=1, \ B_1=b_1+b_2, \ B_2=b_1 b_2,
\]
$\tilde{D}_{n,m,l_1,l_2,l_3 }$ are some coefficients to be determined and
$\sigma_{l_1,l_2,l_3}(c_2,c_3,c_4)$ are Schur polynomials
\footnote{Other bases of symmetric functions, e.g.
$(c_2+c_3+c_4)^i(c_2c_3+c_3c_4+c_4c_1)^j(c_2c_3c_4)^k$ are possible.
However, our choice of Schur polynomials turns out to give simple
expression (\ref{genl}).   } defined as
\eq
\label{e.schurdef}
\sigma_{n_1,n_2,n_3}(x_1,x_2,x_3)=
\left\vert\begin{array}{ccc}
        x_1^{n_1+2}         & x_2^{n_1+2}            & x_3^{n_1+2}      \\
        x_1^{n_2+1}         & x_2^{n_2+1}           & x_3^{n_2+1}        \\
        x_1^{n_3}         & x_2^{n_3}           & x_3^{n_3}
 \end{array}\right\vert / \left\vert\begin{array}{ccc}
        x_1^2         & x_2^2            & x_3^2             \\
        x_1         & x_2           & x_3        \\
        1         & 1           & 1
 \end{array}\right\vert.
 \eqx 
The coefficients can be obtained by comparing the
Taylor expansion of (\ref{shur}) with
$Z_{L=2}^{1/16th, cut}(a_2,b_1,b_2,c_2,c_3,c_4)$. 

The analysis for $L=3,4$ ( with the sum over $l_1$, $l_2$, $l_3$
in (\ref{shur}) from $0$ to $L$ ) is analogous. In this manner we
obtain a fairly simple rational generating functions for $L=2,3,4$.

Given those three functions, it was possible to guess the
partition function for arbitrary $L$.
The final result turns out to have a particularly simple expression in
terms of Schur polynomials namely

\begin{equation}
Z_L^{1/16th}(a_2,b_1,b_2,c_2,c_3,c_4)=\frac{P}{(1-a_2b_1)(1-a_2b_2)},
\label{genl}
\end{equation}
\[
P=\sigma_{L, L, 0} + a_2 \sigma_{L,L - 1,0} + a_2^2 \sigma_{L - 1,L - 1,0} 
\]
\[
+(b_1 + b_2) \left( \sigma_{L,L,1} + a_2 \sigma_{L,L - 1,1} + a_2^2
\sigma_{L - 1,L - 1,1} \right) 
\]
\[
 +b_1 b_2 \left( \sigma_{L,L,2} + a_2 \sigma_{L,L - 1,2} +  a_2^2
 \sigma_{L - 1,L - 1,2} \right),
\]
where $\sigma_{n_1,n_2,n_3}=\sigma_{n_1,n_2,n_3}(c_2,c_3,c_4)$ is the
Schur polynomial (\ref{e.schurdef}).

In order to test the above result further, we performed the analysis for
for $L=5$  with $0 \le n_{a_2}, \ n_{b_1}, \
n_{b_2} \le 2$ and $ 0 \le n_{c_2}, \ n_{c_3}, \ n_{c_4} \le 5$. There are $3^3
6^3=5832$ such configurations and  $414$ which satisfy the central
charge constraint. The corresponding generating function indeed
confirms (\ref{genl}). 

In the remaining part of this section we perform other checks of (\ref{genl}).

\subsection*{The partition function for $\f{1}{8}$BPS states}

 The $\f{1}{8}$BPS states
are obtained by imposing the additional condition $J_1=0$ on the \bps
states \cite{MM}. This is equivalent to
setting the corresponding constraint on the numbers of quanta, namely
\eq
n_{a_2}=0.
\eqx
The central charge condition in this case
\eq
2L+n_{b_1}+n_{b_2}=n_{c_2}+n_{c_3}+n_{c_4},
\eqx
ensures that for given $L$ there is only a finite number of such
configurations.
It follows that the corresponding generating function is a
polynomial in the variables $b_1$, $b_2$, $c_2$, $c_3$, $c_4$. Since
$n_{c_2},n_{c_3},n_{c_4} \le L$ the numbers $n_{b_1}$, $n_{b_2}$ are
also bounded, i.e. $n_{b_1},n_{b_2} \le L$.

These simplifications allow us to perform explicit computations of
the generating function for these states for higher $L$ (we did it for
$L=6$ and $L=7$) and check these results with the general partition
function obtained in the previous section.
Indeed, we find that the resulting functions coincide with (\ref{genl}) after
setting $a_2=0$, i.e. they are
\[
Z_L^{1/8th}(b_1,b_2,c_2,c_3,c_4)=Z_L^{1/16th}(0,b_1,b_2,c_2,c_3,c_4)
\]
\[
=\sigma_{L, L, 0}(c_2,c_3,c_4)  +(b_1 + b_2)\sigma_{L,L,1}(c_2,c_3,c_4)
+ b_1 b_2 \sigma_{L,L,2}(c_2,c_3,c_4).
\]

\subsection*{Checks for operators with many derivatives}

One puzzling feature of the generating function (\ref{genl}) is that
the denominators contain only {\em two} factors: $(1-a_2
b_1)(1-a_2b_2)$. This suggests that the derivatives in \bps states
are essentially commutative (we will discuss this point further in
section 9) which has a crucial impact on the singularity structure of
the \bps partition function.
The test of $\f{1}{8}$BPS states checks the numerator and does not
involve any derivative terms. In order to check for derivatives we
looked at the following configurations for $L=5$ and
$n_{c_3}=n_{c_4}=5$ : i) 7 and 10 $a_2b_1$ derivatives, i.e.
[0,7,7,0,0,5,5,5], [0,10,10,0,0,5,5,5]; ii) 6 derivatives of both
types, i.e. [0,6,6,0,0,5,5,5], [0,6,5,1,0,5,5,5], \ldots
[0,6,3,3,0,5,5,5], iii) states with derivatives and an additional
$c_2$ oscillator, i.e. [0,5,5,0,1,4,5,5], [0,5,4,1,1,4,5,5]. In all
cases, despite the large number of derivatives we found only a
single \bps state in those sectors, which is consistent with (\ref{genl}).

\subsection*{Letter partition function}

Finally, let us note that, although we guessed the partition function
for \bps states starting from $L=2$ and proceeding to $L>2$,
substituting $L=1$ in (\ref{genl}), we recover the letter 
partition function (\ref{e.letter}) which correspond to \bps operators
in a $U(N)$ gauge theory. This is another consistency check of our
analytical formula (\ref{genl}).

\section{Comparision with supergravity}

The one loop \bps partition function can be calculated from the length $L$
partition functions obtained in the previous section in two
steps. First, the single trace partition function is obtained through
\eq
\label{e.zstdef}
Z_{s.t.}=\sum_{L=1}^\infty Z_L^{1/16th},
\eqx
where we sum from $L=1$ since we are considering the partition
function in a $U(N)$ gauge theory\footnote{This will turn out to be
  crucial for comparison with the supergraviton partition
  functions.}. Then, the full \bps partition function is obtained by
passing to multitrace operators through
\eq
\label{e.multi}
Z=\exp \left(\sum_{n=1}^\infty
\frac{1}{n} \left\{ Z^B_{s.t.}(x^n,\ldots)+(-1)^{n+1} 
Z^F_{s.t.}(x^n,\ldots) \right\}\right).
\eqx
In this step we are using the fact that only the planar
one loop dilatation operator is considered. So we are in the strict $N\to \infty$
limit.

The first sum (\ref{e.zstdef}) can be carried out analytically, and
the result is
\eq
\label{e.zst}
Z_{s.t.}=\f{bosons+fermions}{denominator},
\eqx
where
\eqn
denominator \!\!\! &=&\!\!\! (1-a_2 b_1)(1-a_2 b_2)(1-c_2c_3) (1-c_2
c_4)(1-c_3c_4) \nonumber,\\ 
bosons \!\!\!&=& \!\!\! a_2^2 + c_2 c_3 + (c_2 + c_3) (1 + (a_2(b_1 +
b_2)-1) c_2 c_3)  c_4  \nonumber \\ && 
+ c_2 c_3 (a_2 b_1 + a_2 b_2 - 1 + (1 + b_1 b_2 + a_2^2 b_1 b_2 -a_2
(b_1 + b_2)) c_2 c_3) c_4^2 \nonumber,\\ 
fermions\!\!\! &=&\!\!\! (b_1 + b_2) c_2 c_3 c_4 + a_2^2 (b_1 +
b_2)c_2 c_3 c_4  \nonumber\\ && 
+a_2 (c_3 + c_4 + b_1 b_2 c_2^2 c_3 c_4 (c_3 + c_4) + c_2 (c_3 c_4-1)
(b_1b_2 c_3 c_4-1)) \nonumber. 
\eqnx
Let us now compare this result with the single particle
supergraviton \bps partition function (\ref{e.sugra}). In order to
make the comparison possible we express the variables $x$, $v$, $w$, $z$
in terms of $a_2$, $c_2$, $c_3$, $c_4$ and $b_1$, $b_2$ in
terms of $y$. Using the definitions (\ref{e.chargedef}) and
(\ref{e.h0}) we find that $b_1=1/y$, $b_2=y$ and we obtain the
dictionary \eqn
x &=& c_2^{\f{1}{3}} c_3^{\f{1}{3}} c_4^{\f{1}{3}} ,\\
v &=& c_2^{-\f{2}{3}} c_3^{\f{1}{3}} c_4^{\f{1}{3}} ,\\
w &=& c_2^{-\f{1}{3}} c_3^{-\f{1}{3}} c_4^{\f{2}{3}} ,\\
z &=& a_2 c_2^{-\f{2}{3}} c_3^{-\f{2}{3}} c_4^{-\f{2}{3}}.
\eqnx
Remarkably enough, with these substitutions the supergraviton \bps
partition function (\ref{e.sugra}) coincides {\em exactly} with the
one loop single trace \bps partition function (\ref{e.zst}). Thus, the full
\bps partition functions coincide also, since the summation over
multitrace operators is mathematically equivalent to summation over
the supergraviton Fock spaces (c.f. (\ref{e.sugramulti}) and
(\ref{e.multi})).

As a byproduct we note that the resulting $\f{1}{8}$BPS partition
function obtained from the one loop perturbative dilatation operator
exactly coincides with the gauge theory result obtained from the
chiral ring reasoning as in \cite{MM}.

\section{Discussion of large $N$ asymptotics}

Ultimately we are interested in the behavior of the partition
function of \bps states which scales like $\log Z \propto N^2$ and
which therefore can account for the entropy of \bps charged black
holes in $AdS_5 \times S^5$. In \cite{MM} a calculation in the free
theory showed that, for values of the chemical potentials below a
certain value, the partition function has a $N \to \infty$ limit,
while above that value one obtains $\log Z \propto N^2$ scaling.
This analysis follows from formula (\ref{e.fixednfree}) which is
exact for any $N$.

At one loop, we do not have a similar exact formula since the \bps
states are very specific and form just a tiny fraction of all
operators made from the `letters' (\ref{e.letter}). It is thus
interesting to understand whether staying within the $N=\infty$ phase
one can see the transition to the `black hole' phase. Firstly, at
finite $N$, the number of states is
diminished due to trace identities following from the Cayley-Hamilton
theorem. Thus, there is a chance of observing $\log Z \propto N^2$ at a
certain 
value of the parameters only when the corresponding $N=\infty$
partition function has a singularity there or is divergent.

Quite remarkably, our single trace partition function is finite for
all arguments less than 1. This is in stark contrast with the free
\bps partition function which blows up much earlier (see section 5
in \cite{MM}). Let us note that there, this conclusion was reached
from the exact formula (\ref{e.fixednfree}). However one can see
this behavior studying directly the single trace partition function
(\ref{e.zstfree}). We checked that calculating (\ref{e.zstfree}) as
a power series even for the simplest case of {\em two}
noncommutative letters, and studying its radius of convergence
recovers exactly the leading singularity (strictly below 1) which
coincides with the transition point obtained from
(\ref{e.fixednfree}). This experiment gives us confidence that the
knowledge of $N=\infty$ partition function can be a reliable guide
to the singularity structure and hence to the position of the phase
transition.

In order to obtain some rough idea about the structure of the one loop
\bps states we tried to see whether one can introduce some
`effective letters' which would then reproduce the one loop single
trace partition function (\ref{e.zst}). We define effective letter
functions  $z_B^{eff.}(x)$ and $z_F^{eff.}(x)$ for bosons and
fermions respectively using the formula 

\eq
Z_{s.t.}=-\sum_{n=1}^\infty \f{\phi(n)}{n} \log
\left(1-z_B^{eff.}(x^n)-(-1)^{n+1} z_F^{eff.}(x^n) \right).
\label{effl} \eqx
where on the left hand side we put the generating function of single
trace \bps operators obtained in the present paper.
Here, the most natural choice of
chemical potentials is $a_2=x^3$,
$c_2=c_3=c_4=x$ and $b_1=b_2=1$ (see
\cite{MM,Silva}). Expanding the l.h.s and r.h.s of
(\ref{effl}) 
it is possible to solve nonlinear equations for the coefficients of
the Taylor expansion of  $z_B^{eff.}(x)$ and $z_F^{eff.}(x)$.  
We have found unique solutions up to the 28th order of $x$. All the
coefficients are integers however they 
become very large and negative which indicates that any
`noncommutative' letters drastically overcount the \bps 
states.

The above experiment suggests that the building blocks of \bps states
are predominantly commutative similarly to the building blocks of
$\f{1}{8}$BPS states as discussed in section 6.1 of \cite{MM}. This is
further supported by the structure of the denominator in (\ref{e.zst})
which essentially means that only the total number of blocks like
$a_2b_1$ etc. matters -- and not their ordering.

Using the above guiding principle of commutativity we tried to apply
the `plethystic' formalism of \cite{bofeng}. In this formalism, the
partition function at finite $N$ can be reconstructed from the
$N=\infty$ one in the following manner. Suppose that the single trace
bosonic and fermionic partition functions at $N=\infty$ are given by
\eq
Z_{s.t.}^B=\sum_{n=0}^\infty a_n x^n,  \qqqq Z_{s.t.}^F
=\sum_{n=0}^\infty b_n x^n, 
\eqx
then the finite $N$ partition function $Z_N(x)$ is obtained from the
infinite product expansion:
\eq
\f{\prod_{n=1}^\infty (1+g x^n)^{b_n}}{\prod_{n=1}^\infty (1-g
  x^n)^{a_n}} =\sum_{N} Z_N(x) g^N,
\eqx
which in fact exactly reproduces the finite $N$ structure of
$\f{1}{8}$BPS states obtained from chiral ring arguments taking as
input only the $N=\infty$ result. However we
do not see any chance of a $\log Z \propto N^2$ behavior when we apply
this formalism to our partition function.

It has been suggested in the literature \cite{det} that the dual states
contributing mainly to the black hole entropy would be of a
determinant type. For fixed $N$, states which are
determinants of some matrices can be expressed as combinations of
multitrace operators. So at least formally, these states are within
the space of states that we consider (which consists of all multitrace
operators). 

Our conclusion is that in order to see the $\log Z \propto N^2$
scaling, one has to use the whole nonplanar one loop dilatation operator
the properties of which probably have a huge impact on the counting of
\bps states with very many traces.

\section{Summary}

In this paper we determined the partition function of \bps operators
in planar perturbative $\nn=4$ SYM at one loop. We used the oscillator
representation of gauge theory operators and of the planar dilatation
operator. 
In order to obtain the partition function we determined the number of
\bps operators for a certain set of restrictions on the number of
oscillators and for operators of lengths less than 5. Then, we reconstructed a
generating function (assuming that it is a rational function) which
reproduced all these results. Subsequently we made numerous further
checks by evaluating the dilatation operator for higher length and
larger number of (some) oscillators and checking the result with the
proposed generating function.

The main result that we found is an {\em exact} agreement with the
partition function of \bps supergravitons in $AdS_5 \times
S^5$. Consequently we also reproduce exactly, using the one loop
perturbative dilatation operator, the counting of
$\f{1}{8}$BPS states which was previously done on the gauge theory
side using chiral ring reasonings \cite{MM}.

Using the identification of single particle supergraviton states in
terms of short representations of $psu(2,2|4)$ (see \cite{MM}) and the
equality with the gauge theory partition function extracted in the
present work we may identify all \bps states as descendents of $\tr
Z^L$ operators. Thus these states will also persist to be \bps at
higher loop orders. In the process of extracting the partition
function from the 1-loop hamiltonian data we did not use in any way
any information about the representation theory of $psu(2,2|4)$. The
fact that we recover all states in these multiplets is a further check
of this procedure.
 
However it is perhaps a bit surprising, in view of the applications to
black hole entropy, that we do not observe any other 
new primary states (or their descendants). As a word of caution we
note that these might in principle appear for higher lengths and
oscillator occupancy numbers than we could check. However, given the
various checks and consistency with $\f{1}{8}$BPS and extrapolation to
$L=1$ letters we do not think that this is very probable.    

The huge reduction of the number of \bps states with respect to the
free theory reinstated agreement with supergraviton partition function.
However, the transition to a phase with black hole like scaling which
was seen at zero coupling seems to disappear. The form of our
partition function suggests that the constituents generating the \bps
states behave much more like commutative objects than `noncommutative
letters'. We speculate that in order to see the black hole phase
explicitly from gauge theory, one has to use the complete nonplanar
dilatation operator.

\bigskip

\noindent{\bf Acknowledgments.} RJ thanks Juan Maldacena for pointing
out this problem and Niels Obers, Javier Mas and Adam
Rej for interesting discussions. This
work has been supported in part by Polish Ministry of Science and
Information Technologies grant 1P03B04029 (2005-2008), RTN network
ENRAGE MRTN-CT-2004-005616, the Marie Curie ToK COCOS (contract
MTKD-CT-2004-517186) (RJ) and by the Marie Curie Research Training
Network ENIGMA (contract MRNT-CT-2004-5652) (MT). We would like to
thank the Isaac Newton Institute for hospitality when this work was being
finished.


\begin{thebibliography}{99}

\bibitem{adscft}
J.~M.~Maldacena,
``The large N limit of superconformal field theories and supergravity,''
Adv.\ Theor.\ Math.\ Phys.\  {\bf 2} (1998) 231
[Int.\ J.\ Theor.\ Phys.\  {\bf 38} (1999) 1113], [hep-th/9711200];\\
S.~S.~Gubser, I.~R.~Klebanov and A.~M.~Polyakov,
``Gauge theory correlators from non-critical string theory,''
Phys.\ Lett.\ B {\bf 428} (1998) 105, [hep-th/9802109];\\
E.~Witten,
``Anti-de Sitter space and holography,''
Adv.\ Theor.\ Math.\ Phys.\  {\bf 2} (1998) 253, [hep-th/9802150].

\bibitem{berenstein}
  D.~Berenstein,
  ``Large N BPS states and emergent quantum gravity,''
  JHEP {\bf 0601} (2006) 125
  [arXiv:hep-th/0507203].

\bibitem{mandal}
  G.~Mandal,
  ``Fermions from half-BPS supergravity,''
  JHEP {\bf 0508}, 052 (2005)
  [arXiv:hep-th/0502104].

\bibitem{maoz}
  L.~Grant, L.~Maoz, J.~Marsano, K.~Papadodimas and V.~S.~Rychkov,
  ``Minisuperspace quantization of 'bubbling AdS' and free fermion  droplets,''
  JHEP {\bf 0508}, 025 (2005)
  [arXiv:hep-th/0505079].

\bibitem{LLM}
  H.~Lin, O.~Lunin and J.~M.~Maldacena,
  ``Bubbling AdS space and 1/2 BPS geometries,''
  JHEP {\bf 0410}, 025 (2004)
  [arXiv:hep-th/0409174].

\bibitem{other} 
  T.~W.~Brown, P.~J.~Heslop and S.~Ramgoolam,
  ``Diagonal multi-matrix correlators and BPS operators in N=4 SYM,''
  arXiv:0711.0176 [hep-th];\\
  F.~A.~Dolan,
  ``Counting BPS operators in N=4 SYM,''
  Nucl.\ Phys.\  B {\bf 790}, 432 (2008)
  [arXiv:0704.1038 [hep-th]];\\
  M.~Bianchi, F.~A.~Dolan, P.~J.~Heslop and H.~Osborn,
  ``N = 4 superconformal characters and partition functions,''
  Nucl.\ Phys.\  B {\bf 767}, 163 (2007)
  [arXiv:hep-th/0609179].


\bibitem{MM}
  J.~Kinney, J.~M.~Maldacena, S.~Minwalla and S.~Raju,
  ``An index for 4 dimensional super conformal theories,''
  Commun.\ Math.\ Phys.\  {\bf 275}, 209 (2007)
  [arXiv:hep-th/0510251].

\bibitem{bpsbh}
  J.~B.~Gutowski and H.~S.~Reall,
  ``Supersymmetric AdS(5) black holes,''
  JHEP {\bf 0402}, 006 (2004)
  [arXiv:hep-th/0401042];\\
  J.~B.~Gutowski and H.~S.~Reall,
  ``General supersymmetric AdS(5) black holes,''
  JHEP {\bf 0404}, 048 (2004)
  [arXiv:hep-th/0401129];\\
  Z.~W.~Chong, M.~Cvetic, H.~Lu and C.~N.~Pope,
  ``General non-extremal rotating black holes in minimal five-dimensional
  gauged supergravity,''
  Phys.\ Rev.\ Lett.\  {\bf 95}, 161301 (2005)
  [arXiv:hep-th/0506029];\\
  H.~K.~Kunduri, J.~Lucietti and H.~S.~Reall,
  ``Supersymmetric multi-charge AdS(5) black holes,''
  JHEP {\bf 0604} (2006) 036
  [arXiv:hep-th/0601156].


\bibitem{Beisert}
  N.~Beisert,
  ``The complete one-loop dilatation operator of N = 4 super Yang-Mills
  theory,''
  Nucl.\ Phys.\  B {\bf 676}, 3 (2004)
  [arXiv:hep-th/0307015].

\bibitem{Sundborg}
  B.~Sundborg,
  ``The Hagedorn transition, deconfinement and N = 4 SYM theory,''
  Nucl.\ Phys.\  B {\bf 573}, 349 (2000)
  [arXiv:hep-th/9908001].

\bibitem{Aharony}
  O.~Aharony, J.~Marsano, S.~Minwalla, K.~Papadodimas and M.~Van Raamsdonk,
  ``The deconfinement and Hagedorn phase transitions in weakly coupled large N
  gauge theories,''
  Comptes Rendus Physique {\bf 5}, 945 (2004).

\bibitem{Silva}
  P.~J.~Silva,
  ``Thermodynamics at the BPS bound for black holes in AdS,''
  JHEP {\bf 0610}, 022 (2006)
  [arXiv:hep-th/0607056];\\
  P.~J.~Silva,
  ``Phase transitions and statistical mechanics for BPS black holes in
  AdS/CFT,''
  JHEP {\bf 0703}, 015 (2007)
  [arXiv:hep-th/0610163].

\bibitem{bofeng}  
S.~Benvenuti, B.~Feng, A.~Hanany and Y.~H.~He,
  ``Counting BPS operators in gauge theories: Quivers, syzygies and
  plethystics,''
  arXiv:hep-th/0608050;\\
B.~Feng, A.~Hanany and Y.~H.~He,
  ``Counting gauge invariants: The plethystic program,''
  JHEP {\bf 0703}, 090 (2007)
  [arXiv:hep-th/0701063].

\bibitem{det}
  M.~Berkooz, D.~Reichmann and J.~Simon,
  ``A Fermi surface model for large supersymmetric AdS(5) black holes,''
  JHEP {\bf 0701}, 048 (2007)
  [arXiv:hep-th/0604023].


\end{thebibliography}
\end{document}